\documentclass[12pt]{article}
\usepackage{amsmath,amsfonts,amssymb,dcolumn,lscape,subfigure,alltt,authblk}
\usepackage{graphicx,color}
\usepackage{tabularx}
\usepackage{epstopdf}
\graphicspath{{Figures/}}

\begin{document}

\baselineskip=8.0mm
\setlength\parindent{0pt}
\newcommand{\be} {\begin{equation}}
\newcommand{\ee} {\end{equation}}
\newcommand{\Be} {\begin{eqnarray}}
\newcommand{\Ee} {\end{eqnarray}}
\renewcommand{\thefootnote}{\fnsymbol{footnote}}
\def\a{\alpha}
\def\b{\beta}
\def\g{\gamma}
\def\G{\Gamma}
\def\d{\delta}
\def\D{\Delta}
\def\e{\epsilon}
\def\k{\kappa}
\def\l{\lambda}
\def\L{\Lambda}
\def\t{\tau}
\def\om{\omega}
\def\Om{\Omega}
\def\s{\sigma}
\def\lg{\langle}
\def\rg{\rangle}
\author[]{Gregor Diezemann}
\affil[]{Institut f\"ur Physikalische Chemie, Universit\"at Mainz, Duesbergweg 10-14, 55128 Mainz, Germany}
\title
{Nonlinear response theory for Markov processes III: Stochastic models for dipole reorientations}
\maketitle
\begin{abstract}
The nonlinear response of molecular systems undergoing Markovian stochastic reorientations is calculated up to fifth order in the amplitude of the external field.
Time-dependent perturbation theory is used to compute the relevant response functions as in earlier treatments (G. Diezemann, Phys. Rev. E{\bf 85}, 051502 (2012), Phys. Rev. E{\bf 96}, 022150 (2017)).
Here, we consider the reorientational motion of isolated molecules and extend the existing calculations for the model of isotropic rotational diffusion to the model of anisotropic rotational diffusion and to the model of rotational random jumps.
Depending on the values of some model parameters, we observe a hump in the modulus of the nonlinear susceptibility for either of these models.
Interestingly, for the model of rotational random jumps, the appearance of this hump depends on the way the coupling to the external field is modelled in the master equation approach.
If the model of anisotropic rotational diffusion is considered, the orientation of the diffusion tensor relative to the molecular dipole moment and additionally the amount of anisotropy in the rotational diffusion constants determine the detailed shape of the nonlinear response.
In this case, the height of the observed hump is found to increase with increasing 'diffusional anisotropy'.
We discuss our results in relation to the features observed experimentally in supercooled liquids.
\end{abstract}
PACS: 64.70.Q-, 61.20.Lc, 05.40.-a
%
\newpage
\section*{I. Introduction}
The study of the primary relaxation of supercooled liquids by means of dielectric techniques in the linear response regime is standard and allows investigations over an extremely broad frequency range\cite{Kremer02,Lunki:2000,Ranko:2014r}.
Apart from the detailed form of the spectra the nature of the dynamical heterogeneities has been studied using various frequency-selective 
techniques\cite{G23,Ediger00,Israeloff00,Richert02}, including higher-dimensional nuclear magnetic resonance experiments\cite{SRS91,HWZS95,G13} and nonresonant dielectric hole-burning\cite{SBLC96,G16,Chamberlin:2018}.
The latter techniques allow the frequency-selective modification of the spectrum via the application of strong electric fields.

In the recent past, the nonlinear dielectric response of a number of glass-forming systems has been investigated and the results have been used to extract the length scale of the dynamical heterogeneities or the number of correlated particles $N_{\rm corr}$, 
cf.\cite{Richert:2017, Gadige:2017}.
According to theoretical predictions, the modulus of the nonlinear susceptibilities (cubic and higher-order) exhibits a so-called hump, the height of which is directly related to $N_{\rm corr}$\cite{Bouchaud05,Albert:2016}.
It is thus assumed that the origin of the hump is intimately related to the existence of a growing amorphous order or some kind of domain structure.
On the other hand, it has been argued that the reorientational motion of individual molecules gives rise to a monotonous decay of the modulus from a finite low-frequency value to zero at high frequencies and this behavior is found for the model of isotropic rotational diffusion\cite{Albert:2016,Dejardin00,DD95,Kalmykov01}.
It is, however, to be mentioned that also some models lacking spatial correlations like the Box model and variations of this and other phenomenological models have been shown to exhibit a hump in the nonlinear susceptibilities\cite{Richert:2017,Richert06,Ladieu:2012,Buchenau:2017}.

We have computed the third-order and the fifth-order response functions for the well known asymmetric double well potential model of dielectric relaxation and for the simple Gaussian trap model for glassy relaxation and found a hump for certain values of the model 
parameters\cite{G75, G87}.
Additionally, first results for the cubic response for the model of rotational random jumps have been presented in ref.\cite{G92}.

In the analyses of most nonlinear dielectric data it has been assumed that for very low frequencies the heterogeneous nature of the response becomes irrelevant and therefore at these long times the individual reorientational motion of the molecules determines the response.
As mentioned, usually the model of isotropic rotational diffusion has been used to calculate the corresponding resonse functions.
However, it is a well known fact that this model is not able to reproduce a number of aspects related to the noninertial molecular reorientations in supercooled liquids, see e.g.\cite{G42}.
Various models introducing finite angular jumps instead of diffusive motions have been proposed.
One of the first attempts to formulate a general stochastic model of rotational jumps has been provided by Ivanov\cite{Ivanov:1964} and since then a number of different treatments of the rotational motions in liquids have been presented\cite{anders:1972, Bagchi:1991, G31, ALessi:2001}.
Also models explicitly taking into account the dynamical heterogeneities can be used to reproduce a number of the findings related to the reorientational motion in the primary relaxation as monitored with different techniques\cite{G17,G21}.

In the present paper, we present results for the third-order and fifth-order nonlinear response for two models of molecular reorientations.
One model treats the isotropic rotations as random jumps on a sphere and the other model is the one of anisotropic rotational diffusion. 
These models can both be viewed as limiting cases of more general rotational jump models\cite{G31}.

The remainder of the paper is organized as follows. 
In the following chapter, we briefly review the models for molecular reorientations that will be used in the calculations of the nonlinear response afterwards.
After a discussion of the results for the third-order and the fifth-order dynamic susceptibilities we close with some concluding remarks.
\section*{II. Markovian reorientational jump models}
In general, the time-dependent orientation of a molecule is described in terms of the Eulerian angles 
$\Om(t)=(\phi(t),\theta(t),\psi(t))$ relating the axes of a molecular fixed frame and some laboratory fixed axes system.
If $P(\Om,t|\Om_0)$ denotes the probability to find an orientation $\Om$ at time $t$ given that it was $\Om_0$ at an earlier time $t=0$, the  
master equation\cite{vkamp81} can be written in the form
\be\label{P.ME}
{\dot P}(\Om,t|\Om_0) = \int\!\!d\Om'W(\Om|\Om')P(\Om',t|\Om_0)-\int\!\!d\Om'W(\Om'|\Om) P(\Om,t|\Om_0)
\ee
with the probability $W(\Om|\Om')$ for a $\Om'\to\Om$-transition. The inital condition is given by $P(\Om,t=0|\Om_0) = \d(\Om-\Om_0)$
and the probability of finding a given orientation is related to $P(\Om,t|\Om_0)$ via 
$p(\Om,t)=\int\!\!d\Om_0P(\Om,t|\Om_0)p(\Om_0)$.
The model of rotational diffusion is recovered in the limit of small rotation angles in which case the master equation corresponds to a Fokker-Planck equation.
In general, the conditional probability can be expressed in terms of Wigner rotation matrix elements
\be\label{P.Dlmn.allg}
P(\Om,t|\Om_0)=\sum_{l,m_1,m_2,n}\left(\frac{2l+1}{8\pi^2}\right)D_{m_1n}^{(l)*}(\Om_0)D_{m_2n}^{(l)}(\Om)F^{(l)}_{m_1m_2}(t)
\ee
with time dependent coefficients $F^{(l)}_{m_1m_2}(t)$ that are solutions of the respective equations.

In ref.\cite{G31}, we have introduced a reorientational jump model where rotational jumps with fixed angular width $\D\Om$ have been considered.
The transition probabilities for this model can be written in the form:
\Be\label{W.kl}
&&W(\Om_1|\Om_2)=
\G_R\sum_{l,m,n}\left(\frac{2l+1}{8\pi^2}\right)D_{mn}^{(l)*}(\Om_1)D_{mn}^{(l)}(\Om_2)\Lambda_{l,m}(\bar\theta,\bar\phi)
\nonumber\\
&&\hspace{-1.2cm}
\mbox{with}\quad
\Lambda_{l,m}(\bar\theta,\bar\phi)=\cos{(2m\bar\phi)}d_{mm}^{(l)}(\bar\theta)
\Ee
Here, $\bar\theta$ and $\bar\phi$ denote the jump angles and $\G_R$ is the jump rate.
In a diagonal approximation, discussed in detail in ref.\cite{G31}, the $F^{(l)}_{mn}(t)$ in eq.(\ref{P.Dlmn.allg}) are given by $F^{(l)}_{mn}(t)=\d_{m,n}F^{(l)}_m(t)$ with
\be\label{Fmt.gen}
F^{(l)}_m(t)=e^{-\G_R\left(1-\Lambda_{l,m}(\bar\theta,\bar\phi)\right)t}
\ee
The exact result for the model of rotational diffusion is recovered for small jump angles and one has
\be\label{F.ARD}
F^{(l)}_{m}(t) = e^{-\{l(l+1)D_X + m^2(D_Z-D_X)\}t}
\ee
where it is assumed that the rotational diffusion coefficients $D_Y$ and $D_X$ are equal, but not necessarily equal to $D_Z$. 
Eq.(\ref{F.ARD}) follows from $\Lambda_{l,m}(\bar\theta,\bar\phi)\simeq 1-\{l(l+1)-m^2\}(\bar\theta/2)^2+m^2(2\bar\phi^2)$ using 
$D_X=\G_R(\bar\theta/2)^2$ and $D_Z=\G_R(2\bar\phi^2)$.
Only for $D_X\!\!=\!\!D_Z$ the second term in the exponential vanishes and the result for isotropic rotational diffusion,
\[
P_{IRD}(\Om,t|\Om_0)=\sum_{l}\left(\frac{2l+1}{8\pi^2}\right)D_{00}^{(l)*}(\Om_0)D_{00}^{(l)}(\Om)e^{-l(l+1)D_Xt}
\]
 is recovered.
The model of rotational random jumps (with rate $\G_{RJ}$) is obtained by averaging over all jump angles and one finds:
\be\label{F.RJ}
F^{(l)}_{m}(t) = e^{-\G_{RJ}t}
\ee
Insertion of this result into eq.(\ref{P.Dlmn.allg}) yields the well known expression
$P_{RJ}(\Om,t|\Om_0)=\frac{1}{8\pi^2}+e^{-\G_{RJ}t}[\d(\Om-\Om_0)-\frac{1}{8\pi^2}]$\cite{Sillescu:1996}.
In this case, the coefficients are not only independent of $m$ but also do not depend on the rank $l$. 

It has to be mentioned that the orientation $\Om(t)$ is the orientation of the tagged molecule at a given time in a laboratory fixed frame. 
For instance in case of anisotropic rotational diffusion, $\Om(t)$ represents the orientation of the coordinate system of the diffusion tensor (D) in the laboratory fixed frame (L).
Experimentally, however, the orientation of the principal axes system (P) of the relevant interaction in the L-system is observed. 
In case of dielectric relaxation, for instance, the P-system is defined by the orientation of the molecular dipole relative to the D-system.
This means that the expectation value of the dipole moment (the response) is written as: 
\be\label{Mu.t}
\lg M(t)\rg=M\lg D_{00}^{(1)}(\Om_{PL}(t))\rg
=M\sum_n D_{0n}^{(1)}(\Om_{PD})\lg D_{n0}^{(1)}(\Om_{DL}(t))\rg
\ee
where we used the fact that $D_{mn}^{(l)}(\Om_{PL})=\sum_{n'} D_{mn'}^{(l)}(\Om_{PD})D_{n'n}^{(l)}(\Om_{DL})$ and that 
$\Om_{PD}$ is a static quantity defined by the geometry of the molecule considered.
Here $M$ denotes the static value of the dipole moment and $M(\Om_{PL})=M\cos{(\Om_{PL})}$.
The expectation value of an orientation-dependent quantity $A(\Om)$ can be expressed in terms of the solution of the master equation
\[
\lg A(\Om(t))\rg=\int\!d\Om\int\!d\Om_0 A(\Om)P(\Om,t|\Om_0)p^{eq}(\Om_0)
\]
where $p^{eq}(\Om_0)$ denotes the equilibrium probability, in our case $p^{eq}(\Om_0)=1/(8\pi^2)$ because all orientations are equally probable.
\section*{III. Nonlinear response theory for Markov processes}
The response of the system to an external $E$ field applied at time $t=0$ and measured by the moment $M(t)$ is given by eq.(\ref{Mu.t}), which in terms of the solution of the master equation in the presence of a field $E$ is written as
\be\label{Mu.t.G}
\chi(t)=\lg M(t)\rg=M\sum_n D_{0n}^{(1)}(\Om_{PD})\int\!d\Om\int\!d\Om_0 D_{n0}^{(1)}(\Om) P^{(E)}(\Om,t|\Om_0)p^{eq}(\Om_0)
\ee
Here, $\Om$ is a shorthand notation for $\Om_{DL}$.
The time-dependent perturbation theory for the conditional probability $P^{(E)}(\Om,t|\Om_0)$ has been discussed in detail in refs.\cite{G75,G87,G92}.
For the models considered in the present paper, it is sufficient to note that the field is coupled to the transition probabilities $W(\Om_1|\Om_2)$ in eq.(\ref{W.kl}) via the orientation dependent dipole moment  $M(\Om)$ in the following way:
\be\label{W.kl.E}
W^{(E)}(\Om_e|\Om_i)=W(\Om_e|\Om_i)e^{-\b E(t)(\mu M(\Om_i)-(1-\mu)M(\Om_e))}
\ee
where $\mu$ is a model parameter that determines how the system couples to the field. 
For $\mu=1$, the coupling takes place via the initial orientation of the transition ($\Om_i$) and for $\mu=0$ only the destination orientation ($\Om_e$) is relevant. The particular choice $\mu=1/2$ is important for small step reorientations because in this case $(M(\Om_i)-M(\Om_e))\sim dM/d\Om$ and therefore represents the force acting on the system via the application of the field.

As detailed in ref.\cite{G75,G87}, eq.(\ref{W.kl.E}) is used in the master equation for $P^{(E)}(\Om,t|\Om_0)$ and the time-dependent perturbation theory is obtained from a series expansion of $W^{(E)}(\Om_e|\Om_i)$.
From this, the conditional probabilities are obtained in any desired order in the field amplitude. 
We will compute the experimentally relevant response to a sinusoidal field of the form $E(t)=E_0\cos{(\om t)}$.
In the following, we will write for the corresponding susceptibilities monitored after the decay of initial transients:
\Be\label{Chi.om.def}
\chi^{(1)}(t)
&&\hspace{-0.6cm}=
{E_0\over2}\left[e^{-i\om t}\chi_1(\om)+c.c.\right]
\nonumber\\
\chi^{(3)}(t)
&&\hspace{-0.6cm}=
{E_0^3\over2}\left[e^{-i\om t}\chi_3^{(1)}(\om)+e^{-i3\om t}\chi_3^{(3)}(\om)+c.c.\right]
\\
\chi^{(5)}(t)
&&\hspace{-0.6cm}=
{E_0^5\over2}\left[e^{-i\om t}\chi_5^{(1)}(\om)+e^{-i3\om t}\chi_5^{(3)}(\om)+e^{-i5\om t}\chi_5^{(5)}(\om)+c.c.\right]
\nonumber
\Ee
where $c.c.$ denotes the complex conjugate.

In the present paper, we will concentrate on the discussion of the response functions $\chi_k^{(k)}(\om)$, i.e. we focus on the highest frequency component in a given order (third order or fifth order).

We note that we only consider systems that are in thermal equilibrium and therefore the well-known fluctuation-dissipation theorem (FDT) relating the linear response to a short field pulse, $R^{(1)}(t)$, and the autocorrelation function of the dipole moment holds, see e.g.\cite{Chandler87}.
\section*{IV. Results for simple models of molecular reorientations}
Here, we consider the Brownian rotational motion of molecules in terms of the simple models of rotational random jumps (RJ) and anisotropic rotational diffusion (ARD) in addition to the model of isotropic rotational diffusion (IRD).  
In the terminology of ref.\cite{CrausteThibierge10,Brun11} these models describe the 'trivial' dynamics of individual molecules in a glass-forming liquid without any so-called glassy correlations and therefore cannot account for the nontrivial features like the observed hump in the nonlinear response. 
We do not go into technical details of the calculations, which are lengthy but straightforward.
The results for $\chi_k^{(k)}(\om)$, $k$=1,3,5, for the three models considered in the present paper are explicitly given in the Appendix.

The linear dielectric susceptibility for the IRD model and the RJ model can be written in the form:
\be\label{Chi1.iso}
\chi_{1,Z}(\om)=\D\chi_1{1\over1-i\om\t_{10}}\quad\mbox{with}\quad \D\chi_1=\b{M^2\over3}
\ee
where the relaxation time is $\t_{10}=1/(2D_X)$ if $Z$=IRD and $\t_{10}=1/\G_{RJ}$ for $Z$=RJ.
Furthermore, $\b=1/(k_BT)$ denotes the inverse temperature and we will set the Boltzmann constant $k_B$ to unity in all following expressions.
The static linear response (corresponding to $\D\e$ in the dielectric terminology) is denoted by $\D\chi_1$.
For the ARD model, one finds:
\be\label{Chi1.ard}
\chi_{1,ARD}(\om)=\D\chi_1\left({\cos^2{(\Theta)}\over1-i\om\t_{10}}+{\sin^2{(\Theta)}\over1-i\om\t_{11}}\right)
\ee
with $1/\t_{1m}=2D_X+m^2(D_Z-D_X)$ and $\Theta\equiv\Theta_{PD}$ denoting the angle between the z-axes of the molecular axis system (P-system) and of the principal axis system of the diffusion tensor (D-system).
Thus, in this case one has a superposition of two Lorentzians with weights depending on the value of $\Theta$.
Only for $\Theta=0,\pi/2$ one is left with one Lorentzian.
However, the fact that the spectrum is given by a superposition of two Lorentzians is not relevant in the present context because 
in supercooled liquids usually distributions of relaxation times exist that give rise to very broad spectra.
In the general case, the relaxation time is most meaningful defined via the decay time of the normalized dipole autocorrelation function,
\be\label{tau1.def}
\t_1=\int_0^\infty\!\! dt \lg M(t)M(0)\rg_n=\cos^2{(\Theta)}\t_{10}+\sin^2{(\Theta)}\t_{11}
\ee
which reduces to $\t_1\equiv\t_{10}$ for the IRD model and the RJ model.
In the following, we will present all spectra as a function of $\om\t_1$ with the consequence that $\chi_1(\om)$ for the RJ model and IRD model coincide. Furthermore, for the ARD model the spectra for $\Theta=0$ and $\Theta=90^o$ are identical to the one for the IRD model.

In the past, experimental results of nonlinear dielectric spectra have either been presented in terms of real and imaginary part of the susceptibility or, alternatively, the modulus and the phase have been considered.
In particular, it has proven meaningful to scale the modulus by the squared static linear response in the following way:
\be\label{X35.def}
X_3(\om)={T\over(\D\chi_1)^2}\left|\chi_3^{(3)}(\om)\right|
\quad\mbox{and}\quad
X_5(\om)={T^2\over(\D\chi_1)^3}\left|\chi_5^{(5)}(\om)\right|
\ee
These definitions eliminate the trivial temperature dependences, $\chi_3^{(3)}\propto\b^3$ and $\chi_5^{(5)}\propto\b^5$.
Using $\D\chi_1=\b M^2/3$  and the expressions given in the Appendix, one can write the moduli in terms of the spectral functions for each of the models considered:
\be\label{X35.rots}
X_{3,Z}(\om)={3\over20}\left|S_{3,Z}(\om\t_1)\right|
\quad;\quad
X_{5,Z}(\om)={9\over560}\left|S_{5,Z}(\om\t_1)\right|
\ee
where $Z$ is an abbreviation for IRD, ARD or RJ.
The low-frequency limits for all models considered coincide and are given by:
\be\label{X35.limits}
X_{3,Z}(\om\to0)={1\over20}
\quad\mbox{and}\quad
X_{5,Z}(\om\to0)={1\over280}
\ee
However, for high frequencies the limiting behavior for the RJ model differs from that for the diffusion models.
One finds $X_{3,Y}(\om\to\infty)\sim\om^{-3}$ and $X_{5,Y}(\om\to\infty)\sim\om^{-5}$ for $Y$=IRD, ARD while for the RJ model one has $X_{k,RJ}(\om\to\infty)\sim\om^{-1}$ for both $X_3$ and $X_5$.
However, we will not further discuss this high frequency behavior as it does not appear to be observable in supercooled liquids due to the existence of other relaxation phenomena such as the so-called wing or secondary processes at higher frequencies\cite{Lunki:2000}.

\begin{figure}[h!]
\centering
\includegraphics[width=15cm]{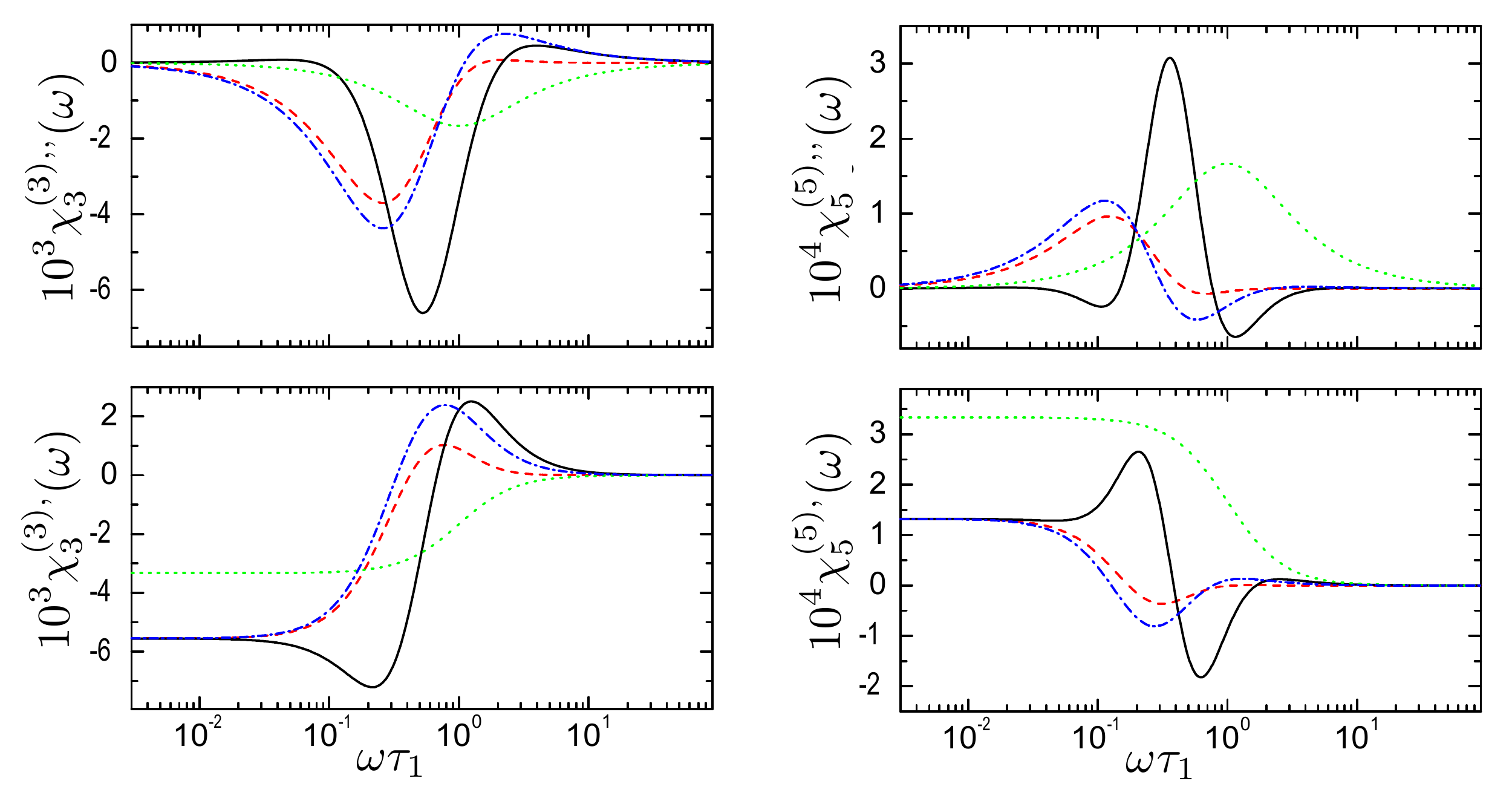}
\vspace{-0.25cm}
\caption{Real and imaginary part of the cubic susceptibility $\chi_3^{(3)}(\om)$ (left) and the fifth-order susceptibility 
$\chi_5^{(5)}(\om)$ for the models of rotational diffusion (red dashed lines) and rotational random jumps for $\mu=1$ (black full lines) and $\mu=0$ (blue dot-dashed lines).
The green dotted lines represent the linear response, cf. eq.(\ref{Chi1.iso}), scaled by a factor of ten.
}
\label{Fig1}
\end{figure}

In Fig.\ref{Fig1} we show the real and the imaginary part of the third-order and the fifth-order response for the model of isotropic rotational diffusion and for the RJ model. 
In the latter case, we used two values for the parameter $\mu$ that describes the coupling to the external field, cf. 
eq.(\ref{W.kl.E}).
For $\mu=1$ (black full lines) the coupling takes place via the initial state of a rotational transition and for $\mu=0$ (blue dot-dashed lines) only the destination state is relevant. 
For the RJ model, it appears that $\mu=1$ is the more natural choice because the idea underlying the model is that every single transition completely decorrelates the orientation in the sense that starting from a given orientation any other can be reached in a single step.
It is then meaningful to assume that starting with a coupling to an initial orientation according to $(-E\cdot M(\Om_i))$, 
an average over all possible destination orientations is to be performed, $\lg(-E\cdot M(\Om_e))\rg$, which vanishes.
Therefore, according to eq.(\ref{W.kl.E}), this corresponds to choosing $\mu=1$.

The overall behavior of the results for the IRD model and the RJ model is quite similar for both response functions, in particular if $\mu=0$ is chosen in case of the RJ model (blue dot-dashed lines).
For $\mu=1$ (black full lines), the deviations from a monotonous behavior of the real parts are stronger.

\begin{figure}[h!]
\centering
\includegraphics[width=16cm]{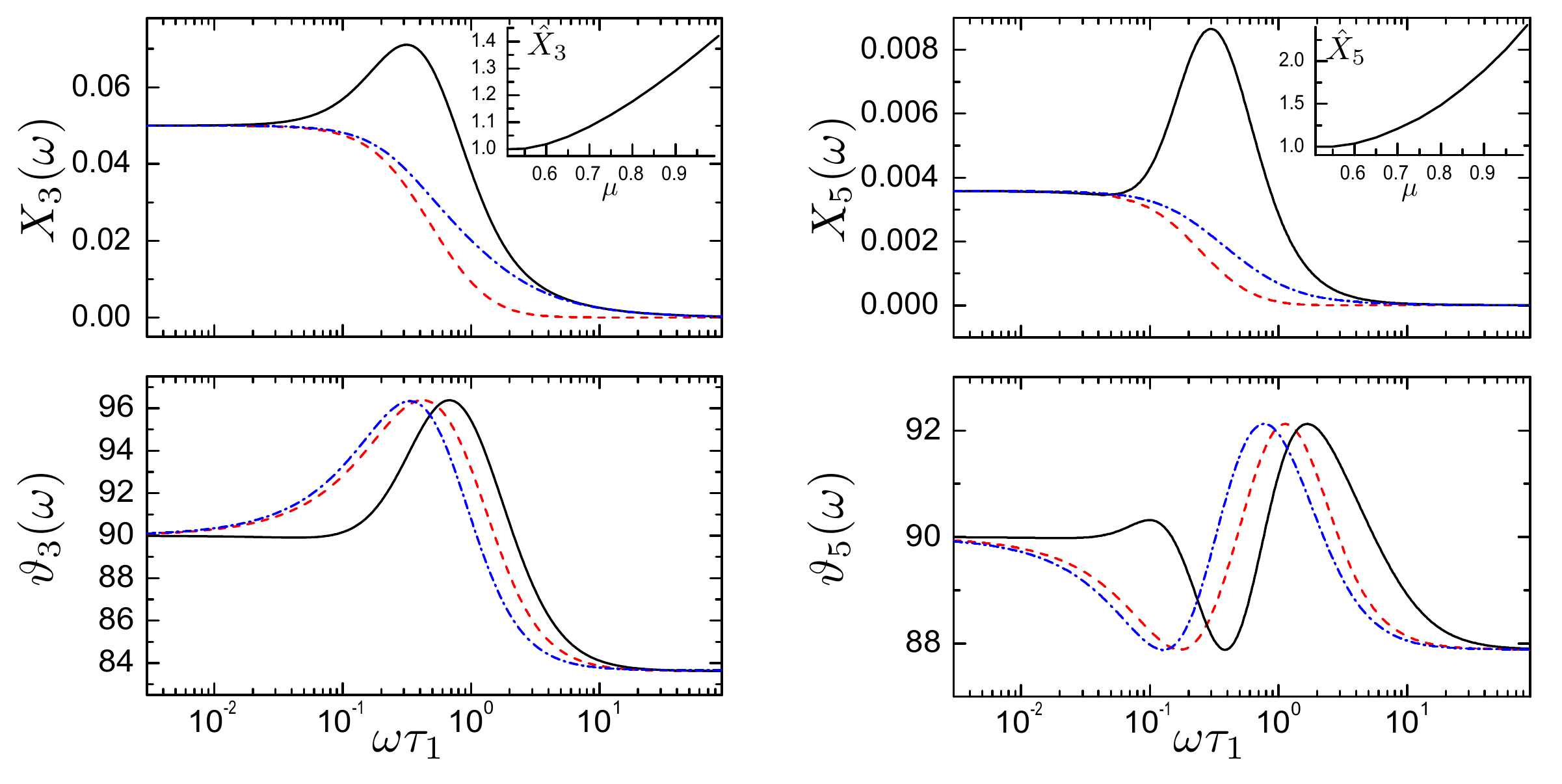}
\vspace{-0.5cm}
\caption{upper panels: $X_3(\om)$ (left) and $X_5(\om)$ (right) for the models of rotational diffusion (red dashed lines) and rotational random jumps (black full lines ($\mu=1$) and blue dot-dashed lines ($\mu=0$)).
The insets show the relative maximum value of the hump, $\hat X_k=X_k^{\rm max}/X_k(0)$, as a function of $\mu$.
Lower panels: Phase $\vartheta_k(\om)=\rm acos{(\chi_k^{(k),,}(\om)/\chi_k^{(k),}(\om))}$ (in deg.) as a function of frequency.
}
\label{Fig2}
\end{figure}

In Fig.\ref{Fig2} the modulus and the phase are shown for the same parameters as in Fig.\ref{Fig1}.
It can be seen that that in both cases, the third-order and the fifth-order response, the RJ model with $\mu=1$ gives rise to a hump in $X_k(\om)$.
The insets in the upper panels of Fig.\ref{Fig2} show the relative magnitude $\hat X_k=X_k^{\rm max}/X_k(0)$ of the hump as a function of the parameter $\mu$.
For values $\mu\lesssim1/2$, no hump appears.
For larger values of $\mu$ one observes a clear hump with a maximum at a frequency somewhat smaller than $\om\t_1\simeq1$.
This means that the simple model of isotropic rotational random jumps in which each transition completely destroys the orientational correlations yields results similar to the nonlinear response observed for supercooled liquids. 
However, the behavior observed in Fig.\ref{Fig2} does not depend on temperature, since all spectral functions are only dependent on the scaled frequency $\om\t_1$, cf. the expressions given in the Appendix. 
Thus, the experimentally observed decrease of the height of the hump with increasing temperature can only be modelled by changing the parameters of the model, in particular the value of $\mu$.

\begin{figure}[h!]
\centering
\includegraphics[width=16cm]{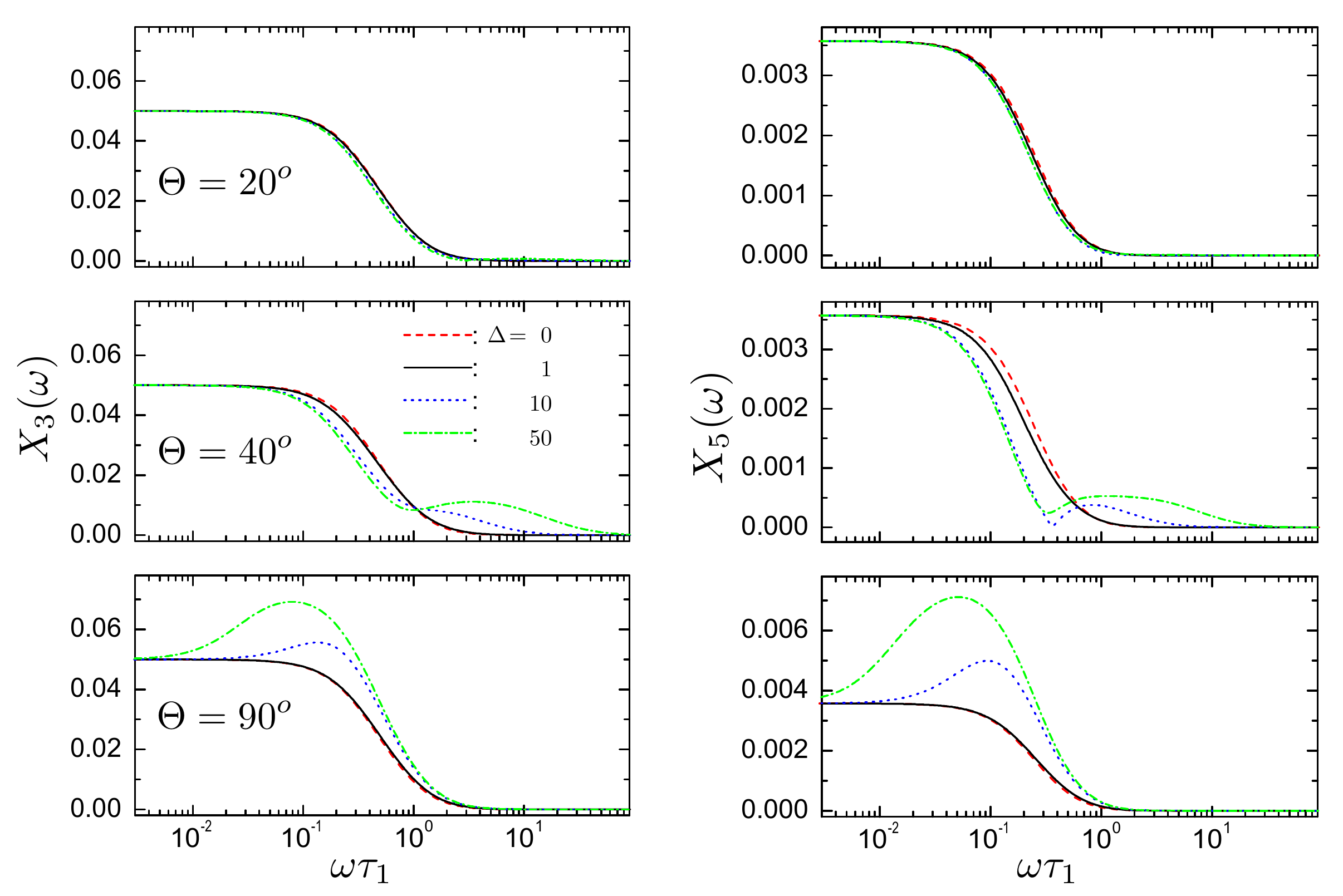}
\vspace{-0.25cm}
\caption{$X_3(\om)$ (left) and $X_5(\om)$ (right) for the models of isotropic rotational diffusion (red dashed lines) and anisotropic rotational diffusion for parameters as indicated in the figure.
Here, $\D=D_Z-D_X$.
}
\label{Fig3}
\end{figure}

In Fig.\ref{Fig3} we show $X_k(\om)$ for the model of anisotropic rotational diffusion for various values of the angle $\Theta$.
For $\Theta=0$, the results are identical to those for the IRD model.
The same holds for vanishing 'diffusional anisotropy', $\D=0$.
For small values of $\Theta$, there are only minor differences between the results for the two models.
With increasing $\Theta$, a shoulder or a peak at higher frequencies develops depending on the value of $\D$.
This is clearly observable for $\Theta=40^o$, where a shoulder is found for $\D=10$ and a peak for $\D=50$.
This behavior with a varying height of the high-frequency peak is observed up to angles of approximately $\Theta=70^o$ (for $\D=10$).
In this regime also the linear susceptibility is composed of two Lorentzian with comparable intensities.
For higher values of $\Theta$ the peak shifts to lower frequencies and a single hump is observed in the moduli $X_3$ and $X_5$ as is most prominently seen for $\Theta=90^o$ in Fig.\ref{Fig3}.
Note that for $\Theta=90^o$, the scaled linear response of the ARD model coincides with the corresponding one for the IRD model.
This does not hold for the nonlinear response functions.
The overall behavior of $X_5(\om)$ is very similar to the one of $X_3(\om)$.
In both cases the position of the hump shifts to slightly smaller frequencies with increasing $\D$ and at the same time it broadens somewhat.

\begin{figure}[h!]
\centering
\includegraphics[width=8cm]{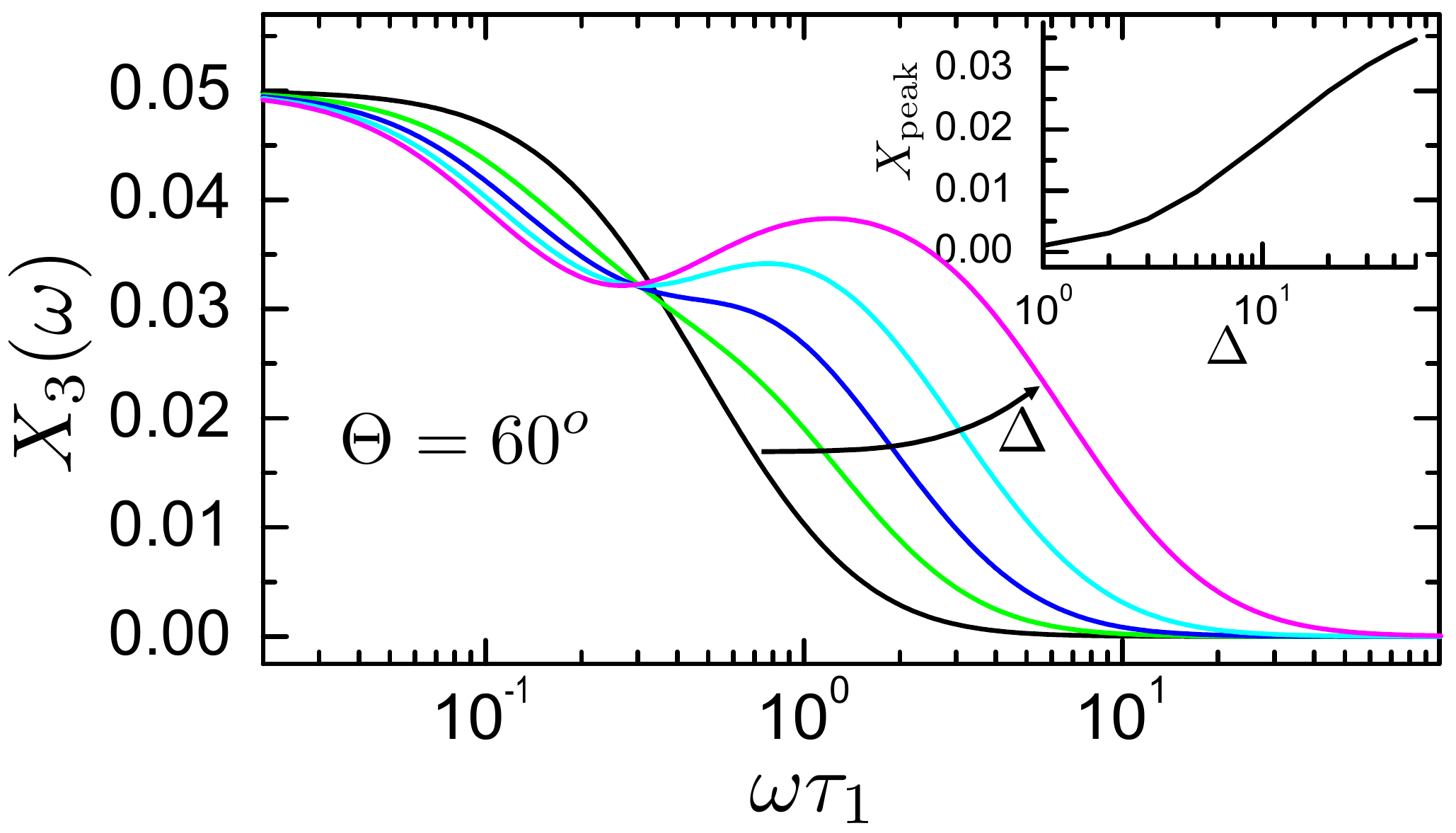}
\vspace{-0.5cm}
\caption{$X_3(\om)$ for the model of anisotropic rotational diffusion for $\Theta=60^o$.
The values of $\D$ are $\D=1,\,5,\,10,\,20,\,50$ in the direction of the arrow.
The inset shows the value of the maximum of the high-frequency peak.
}
\label{Fig4}
\end{figure}

\begin{figure}[h!]
\centering
\includegraphics[width=8cm]{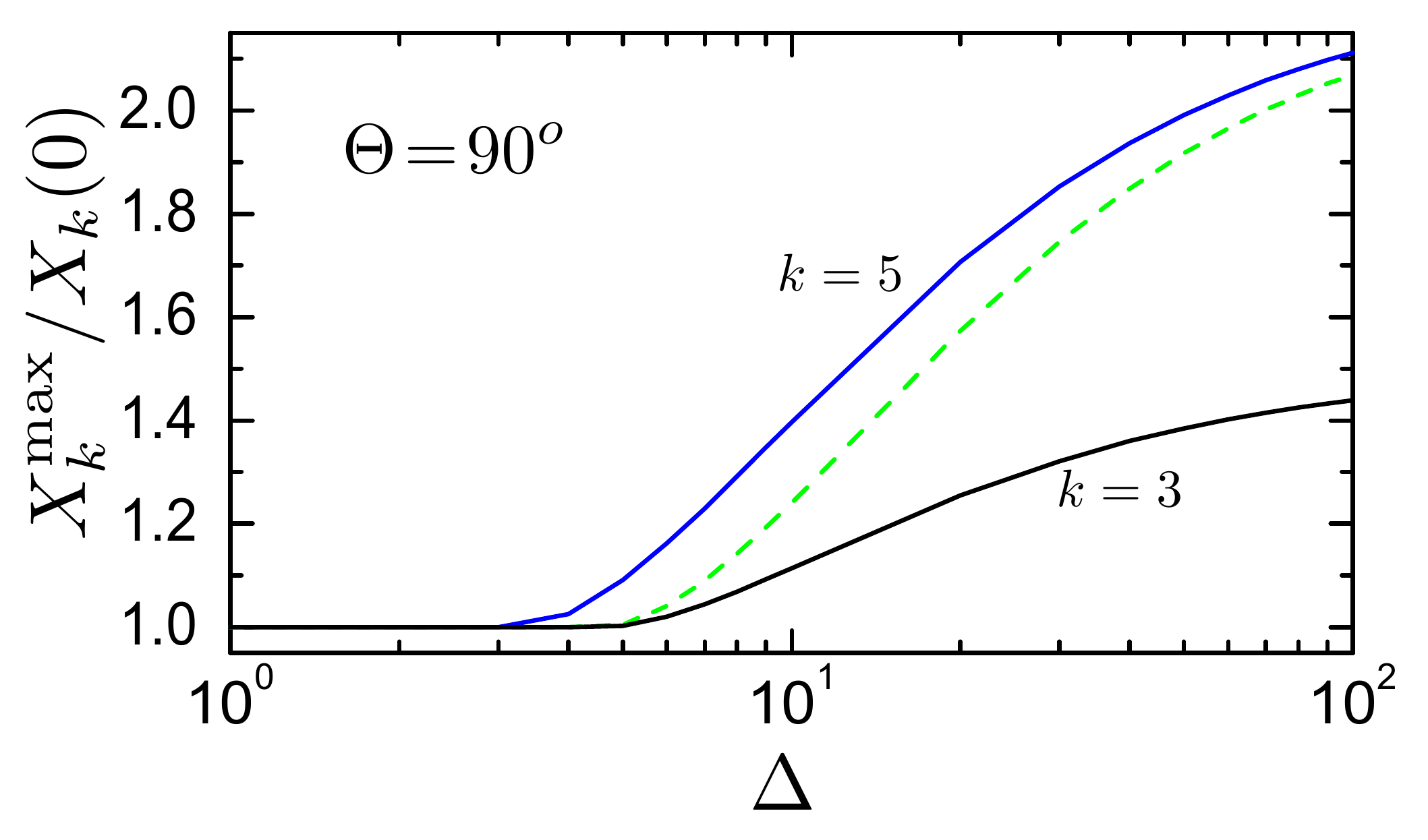}
\vspace{-0.5cm}
\caption{Maximum height scaled to the value at zero frequency, $\hat X_k=X_k^{\rm max}/X_k(0)$ for $\Theta\!=\!90^o$. 
Lower black line: $\hat X_3$, upper blue line: $\hat X_5$ and dashed green line: $(\hat X_3)^2$.
}
\label{Fig5}
\end{figure}

As an example for the behavior at intermediate angles $\Theta$, we plot $X_3(\om)$ for $\Theta=60^o$ and various values of the diffusional anisotropy $\D$ in Fig.\ref{Fig4}.
It is obvious how for smaller values of $\D$ a shoulder evolves that turns into a secondary peak for larger anisotropy.
The inset in Fig.\ref{Fig4} shows the increase of the height of the secondary peak with increasing $\D$.
This behavior is similar to the corresponding one for those values of $\D$ for which a hump is observed.
This fact is detailed in Fig.\ref{Fig5}, where we present the maximum height of the hump for $\Theta=90^o$.
It is apparent that the relative magnitude of $\hat  X_5=X_5^{\rm max}/X_5(0)$ is larger than the corresponding third-order quantity.
In green (dashed line), we show the square of the relative amplitude of $X_3$ as one should approximately have $X_5\sim X_3^2$ under some additional assumptions\cite{Albert:2016}.
As for the model of rotational random jumps, also in the present case the results are independent of temperature.
\section*{V. Conclusions}
While linear susceptibilities are related to equilibrium time correlation functions via the FDT and well known relations among the real and imaginary parts of $\chi_1(\om)$ exist, the situation is different for nonlinear response functions.
In order to learn about the information content of the nonlinear susceptibilities one has to consider the results of explicit model calculations.
As mentioned in the Introduction, a number of such calculations have been performed for different models exhibiting glassy relaxation in the past and in some cases a behavior similar to what is observed in supercooled liquids was obtained.

The interpretations of the hump observed in the nonlinear susceptibilities of glass-forming liquids in terms of growing amorphous order are mainly concerned with the phenomena in the frequency range on the order of the inverse primary relaxation time of the system. 
At longer time scales or smaller frequencies it is assumed that the response does not reflect the dynamic heterogeneities but is due to the rotational motion of individual molecules because exchange processes average out the heterogenous nature of the relaxation.
This is the origin of the mentioned 'trivial' contribution to the nonlinear response functions.
Usually, the model of isotropic rotational diffusion (IRD) is used for the computation of the nonlinear response due to this trivial relaxation.

In the present paper, we have also considered the rotational motion of individual molecules in supercooled liquids.
In addition to the IRD model, we considered two models for Brownian rotational motion, namely the model of isotropic rotational random jumps (RJ) and the model of anisotropic rotational diffusion (ARD). 
These models are considered because it is known for a very long time, that the IRD model does not give a reasonable description of the rotational motion in supercooled liquids and that models using finite jump angles yield more reliable results.
In addition, most molecules showing a significant glass-forming ability are not adequately described as spherical and the rotational motion might show deviations from isotropy.

For both, the RJ model and the ARD model, we find a hump in the moduli of the nonlinear susceptibilities, $X_3$ and $X_5$, for some values of the model parameters.
This means that also models for the 'trivial' contribution to the response can show features similar to what is observed in experiments on supercooled liquids at some temperature.
A temperature dependent height of the hump can only be modelled via changing relevant model parameters.
We only discussed the intrinsic features of the response functions and did not attempt to model the response typically observed in supercooled liquids. 
In order to do so, one would use one of the models for the rotational motion and fit the linear susceptibility using a distribution of relaxation times. The nonlinear response could then be fitted by adjusting the other model parameters like $\mu$ in case of the RJ model and $\Theta$, $\D$ for the ARD model.
On the other hand, if one assumes a different model for instance including cooperativity for the primary relaxation and uses the model for the reorientational motion solely for the trivial contribution, one has to be careful when extracting quantities related to the height of the experimentally observed hump in $X_3$ or $X_5$.

The calculations presented here clearly indicate that it is not straightforward to extract informations from nonlinear response functions. 
As shown earlier, in some cases model calculations can help to discriminate among different models\cite{G87}.
However, general arguments regarding the detailed behavior of nonlinear susceptibilites are rare and more theoretical effort will be required to obtain conclusive results.

\section*{Acknowledgement}
Useful discussions with Roland B\"ohmer, Gerald Hinze, Francois Ladieu and Jeppe Dyre are gratefully acknowledged.

\begin{appendix}
\section*{Appendix A: The model of isotropic rotational diffusion}
\setcounter{equation}{0}
\renewcommand{\theequation}{A.\arabic{equation}}
The linear and nonlinear dielectric spectra for the model of isotropic rotational diffusion have been calculated\cite{Dejardin00, Albert:2016} and the corresponding expressions are repeated here for convenience.
The method used in ref.\cite{Dejardin00} is slightly different from the time-dependent perturbation theory that is used in the present approach.
The results, however, agree up to a constant, which is defined indirectly by the definition of the susceptibilities, 
cf. eq.(\ref{Chi.om.def}).
The linear response is determined by the expression given in eq.(\ref{Chi1.iso}).
For the cubic response, we consider the $3\om$-component, which for this model is given by
\be\label{chi33.IRD}
\chi_{3,IRD}^{(3)}(\om)=-{1\over60}\b^3M^4S_{3,IRD}(\om\t_1)
\;;\;
S_{3,IRD}(x)=\frac{1}{(1-ix)(1-i3x)(3-i2x)}
\ee
Here, $x=\om\t_1$ with $\t_1=1/(2D_R)$.

The $5\om$-component of the fifth-order response is given by:
\Be\label{chi55.IRD}
\chi_{5,IRD}^{(5)}(\om)=&&\hspace{-0.6cm}
{1\over1680}\b^5M^6S_{5,IRD}(\om\t_1)
\nonumber
\\
S_{5,IRD}(x)=&&\hspace{-0.6cm}
\frac{4-i5x}{(1-i5x)(3-i4x)(1-i3x)(3-i2x)(1-ix)(2-ix)}
\Ee
\section*{Appendix B: The model of ansotropic rotational diffusion}
\setcounter{equation}{0}
\renewcommand{\theequation}{B.\arabic{equation}}
For anisotropic rotational diffusion, we proceed in the following way.
We start the calculation from the rotational jump model discussed in Section II and perform the limit of small jump angles in the end of the calculations. Furthermore, we use $\mu=1/2$ as this value is the relevant one in the diffusive limit of the master equation.
Technically, this means that in the time dependent perturbation expansion of the propagator ${\bf G}^{(E)}$, the matrix of 
$P^{(E)}(\Om,t|\Om_0)$, only the terms containing the linear perturbation ${\cal V}^{(1)}$ have to be considered in eq.(8) of 
ref.\cite{G87} because all other terms vanish in the diffusive limit.
In the notation used there, this can be written as ${\bf G}^{(n)}={\bf G}\otimes\left[{\cal V}^{(1)}\otimes{\bf G}\right]^n$ with 
$n=3$ or $n=5$. The symbol $\otimes$ indicates the convolution 
${\bf G}\otimes{\cal V}^{(1)}\otimes{\bf G}\equiv\int_{t_0}^t\!dt'{\bf G}(t,t'){\cal V}^{(1)}(t'){\bf G}(t',t_0)$.

As mentioned in the text, in the case of anisotropic reorientational motions, the results do not only depend on the overall value of the molecular  dipole moment, $M$, but also on the orientation of the diffusion tensor relative to the applied electric field, cf. eq.(\ref{Mu.t}), which we write as ($\Om(t)\equiv\Om_{DL}(t)$): 
\be\label{xi.def}
\lg M(t)\rg=M\sum_n \xi_n\lg D_{n0}^{(1)}(\Om(t))\rg 
\quad\mbox{with}\quad \xi_n=D_{0n}^{(1)}(\Om_{PD}).
\ee
Without going into the details of the lengthy calculations, we simply will present the results in a compact form.
We define the following function:
\Be\label{Y.def}
Y(L_1,M_1;L_2,M_2)&=&\sum_N\xi_N(\G_{L_2,M_2}+\G_{1,N}-\G_{L_1,M_1})(-1)^{M_2}\times
\nonumber\\
&&\hspace{0.5cm}\times
\left(2L_2+1\right)
\left(\begin{matrix}
 1 & L_1 & L_2\\
 N & M_1 & -M_2
\end{matrix}\right)
\left(\begin{matrix}
 1 & L_1 & L_2\\
 0 & 0 & 0
\end{matrix}\right)
\Ee
Here,
$\left(\begin{smallmatrix}
 1 & L_1 & L_2\\
 N & M_1 & -M_2
\end{smallmatrix}\right)$ denotes a Wigner 3-j symbol and the rates are defined by:
\be\label{Gam.lm}
\G_{L,M}=L(L+1)D_X+M^2(D_Z-D_X)
\ee
cf. the discussion in the context of eq.(\ref{F.ARD}).

With this, we find for the cubic response:
\be\label{chi33.ARD}
\chi_{3,ARD}^{(3)}(\om)={\b^3M^4\over60}\!\!\!\sum_{m_1,m_2,m_3}\!\!\!(-)^{m_1}\xi_{-m_1}\xi_{m_3}\G_{1,m_3}
Y(1,m_3;2,m_2)Y(2,m_2;1,m_1)G_{m_1,m_2,m_3}(\om)
\ee
with
\be\label{G3.def}
G_{m_1,m_2,m_3}(\om)={5\over4}\left[(\G_{1,m_1}-i3\om)(\G_{2,m_2}-i2\om)(\G_{1,m_3}-i\om)\right]^{-1}
\ee
For the fifth-order response, one has:
\Be\label{chi55.ARD}
\chi_{5,ARD}^{(5)}(\om)={\b^5M^6\over1680}\!\!\!\sum_{L=1,3}\sum_{m_1\cdots m_5}\!\!\!(-)^{m_1}\xi_{-m_1}\xi_{m_5}\G_{1,m_5}
Y(1,m_5;2,m_4)Y(2,m_4;L,m_3)
\!\!\times
\nonumber\\
&&\hspace{-9.5cm}\times
Y(L,m_3;2,m_2)Y(2,m_2;1,m_1)G_{m_1,m_2,m_3,m_4,m_5}(\om)
\Ee
\Be\label{G5.def}
&&G_{m_1,m_2,m_3,m_4,m_5}(\om)=
\nonumber\\
&&\hspace{0.5cm}{35\over16}
\left[(\G_{1,m_1}-i5\om)(\G_{2,m_2}-i4\om)(\G_{L,m_3}-i3\om)(\G_{2,m_4}-i2\om)(\G_{1,m_5}-i\om)\right]^{-1}
\Ee
\section*{Appendix C: The model of isotropic rotational random jumps}
\setcounter{equation}{0}
\renewcommand{\theequation}{C.\arabic{equation}}
In this case, one has to consider a ME and one has to fix the value of $\mu$ in eq.(\ref{W.kl.E}) as discussed in the text.
Also in this case, we do not present details of the lengthy calculations and only present the results.

The linear response is given by the same expression as for the model of rotational diffusion, eq.(\ref{Chi1.iso}), with the replacement $\t_1=1/\Gamma_{RJ}$.
Also the third-order response can be written in a form that is very similar to eq.(\ref{chi33.IRD}). However, the spectral function is quite different and this gives rise to a different behavior.
\be\label{chi33.RJ}
\chi_{3,RJ}^{(3)}(\om)={1\over60}\b^3M^4S_{3,RJ}(\om/\Gamma_{RJ})
\;;\;
S_{3,RJ}(x)={1\over3}\left[I_{3,0}(x)+4\mu I_{3,1}(x)+4\mu^2 I_{3,2}(x)\right]
\ee
with $x=\om/\Gamma_{RJ}$.
Here, the individual terms are given by:
\Be\label{I3.mu}
I_{3,0}(x)=&&\hspace{-0.6cm}
-\frac{2+i3x}{2(1-ix)(1-i3x)}\nonumber
\\
I_{3,1}(x)=&&\hspace{-0.6cm}
\frac{ix}{(1-i2x)(1-i3x)}
\\
I_{3,2}(x)=&&\hspace{-0.6cm}
\frac{-2x^2+ix}{2(1-ix)(1-i2x)(1-i3x)}\nonumber
\Ee
For the fifth-order susceptibility, we find:
\Be\label{chi55.RJ}
\chi_{5,RJ}^{(5)}(\om)=&&\hspace{-0.6cm}
{1\over1680}\b^5M^6S_{5,RJ}(\om/\Gamma_{RJ})
\nonumber
\\
S_{5,RJ}(x)=&&\hspace{-0.6cm}
{1\over9}\left[{1\over8}I_{5,0}(x)-2\mu I_{5,1}(x)+\mu^2 I_{5,2}(x)-4\mu^3 I_{5,3}(x)+\mu^4 I_{5,4}(x)\right]
\Ee
where the spectral functions are given by:
\Be\label{I5.mu}
I_{5,0}(x)=&&\hspace{-0.6cm}
\frac{16-27x^2+i69x}{(1-ix)(1-i3x)(1-i5x)}
\nonumber\\
I_{5,1}(x)=&&\hspace{-0.6cm}
\frac{40x^2+ix(15+36x^2)}{(1-i2x)(1-i3x)(1-i4x)(1-i5x)}
\nonumber\\
I_{5,2}(x)=&&\hspace{-0.6cm}
\frac{59x^2+144x^4-ix(3+128x^2)}{(1-ix)(1-i2x)(1-i3x)(1-i4x)(1-i5x)}
\\
I_{5,3}(x)=&&\hspace{-0.6cm}
\frac{32x^2+36x^4-ix(3+34x^2)}{(1-ix)(1-i2x)(1-i3x)(1-i4x)(1-i5x)}
\nonumber\\
I_{5,4}(x)=&&\hspace{-0.6cm}
\frac{-57x^2+72x^4-ix(4+226x^2)}{(1-ix)(1-i2x)(1-i3x)(1-i4x)(1-i5x)}
\nonumber
\Ee
For convenience, we give the expression for the particular choice $\mu=1$:
\be\label{S3R.mu1}
S_{3,RJ}^{(\mu=1)}(x)=\frac{-2-6x^2+i13x}{6(1-ix)(1-i2x)(1-i3x)}
\ee
and
\be\label{S5R.mu1}
S_{5,RJ}^{(\mu=1)}(x)=\frac{16-1629x^2+216x^4-ix(227+2070x^2)}{72(1-ix)(1-i2x)(1-i3x)(1-i4x)(1-i5x)}
\ee
\end{appendix}
\end{document}